\documentclass{article}
\usepackage{amssymb}
\usepackage{amsmath}
\usepackage{amsfonts}
\usepackage{graphicx}
\usepackage{epstopdf}
\usepackage{listings}
\usepackage{hyperref}
\usepackage{bookmark}
\usepackage{placeins}
\usepackage{color}
\usepackage{float}
\usepackage{enumitem}
\usepackage{tikz}
\usetikzlibrary{arrows.meta}
\usetikzlibrary{bending}
\usepackage{amsthm}

\title{{\bf {On linear resolvability of universal quantum dimensions}\vspace{.2cm}}
\author{{\bf M.Y. Avetisyan and R.L. Mkrtchyan}}
\date{ }
}
\begin{document}
\maketitle
\date{ }
\begin{center}
  {\small {\it Yerevan Physics Institute, Yerevan, Armenia}}\\
\end{center}

\vspace{0.125cm}

\begin{abstract}

In his study of finite (Vassiliev's) knot invariants Vogel introduced the so-called universal parameters, 
belonging to projective plane, which particularly parameterize the simple Lie algebras by the Vogel's table.
 Subsequently a number of quantities, such as some universal knot invariants, (quantum) dimensions of simple Lie 
 algebras, etc., have been represented in terms of these parameters, i.e. in the
  universal form.  
We prove that at the points from the Vogel's table all known universal quantum dimension formulae 
 are linearly resolvable, i.e. yield finite answers even if these points are singular, 
provided one restricts them to the appropriate lines.
We show, that the same phenomenon takes place for another three distinguished points in Vogel's plane - 
$E_{7\frac{1}{2}}, X_1,$ and $X_2$.
We also examine the same formulae on linear resolvability 
at the remaining 48 distinguished points in Vogel's plane,
 which correspond to the so-called $Y$-objects, and
  discover that there are three points among them, which are
  regular for all known quantum dimension formulae.
  Two of them happen to be sharing a remarkable similarity with the simple Lie algebras,
  namely, the universal formulae yield integer-valued outputs (dimensions) at those
   points in the classical limit.
	
\end{abstract}

\section*{Introduction}

The category of the Universal Lie Algebra, which was meant to generalize the simple Lie algebras, was introduced by
 Vogel in his study of the most general finite knot invariants \cite{V0,V}.
 These works are also connected with the hypothesis of Deligne \cite{Del,DM,Cohen} on series of the 
 exceptional Lie algebras.

 One of the outputs of Vogel's works was the introduction of a parameterization of simple Lie algebras by three
 so-called  universal parameters -
  $\alpha, \beta, \gamma$, which are the homogeneous coordinates of the projective (Vogel's) plane, relevant up to rescaling and permutations\footnote{Some details of Vogel's approach, particularly Vogel's table \ref{tab:V2}, in which the corresponding values of the universal parameters for all simple Lie algebras are given, are presented in Appendix A.}.  
	
He particularly presented a number of (universal, by definition) dimension formulae, e.g. expressions 
for dimensions of various irreducible
representations of simple Lie algebras in terms of rational functions of these universal parameters. 
As noted in \cite{Cohen}, universal formulae are actually dealing with algebras, which are semidirect 
products of simple Lie algebras with the automorphisms of their Dynkin diagram. 

Subsequently a number of universal formulae, particularly for dimensions and quantum dimensions, 
have been derived in
\cite{LM1, M16QD, AM, X2kng}, as well as Diophantine classification of simple Lie algebras \cite{RoadM}, 
based on the universal quantum dimension of the adjoint representation (\ref{cad})
 \cite{W,MV}, has been worked out. For example, the quantum dimension of the adjoint
  representation of each of the simple Lie algebra is given by the following function:

 \begin{eqnarray}  \label{cad}
 f(x) &=& -\frac{\text{sinh}\left(\frac{\gamma+2\beta+2\alpha}{4}x\right)}{\text{sinh}\left(\frac{\alpha}{4}x\right)}
 \frac{\text{sinh}\left(\frac{2\gamma+\beta+2\alpha}{4}x\right)}{\text{sinh}\left(\frac{\beta}{4}x\right)}
 \frac{\text{sinh}\left(\frac{2\gamma+2\beta+\alpha}{4}x\right)}{\text{sinh}\left(\frac{\gamma}{4}x\right)}
 \end{eqnarray}

 An important application of universal dimension formulae of representations of simple Lie algebras is the possibility to rewrite the knot polynomials 
(which are not finite Vassiliev invariants) for some classes of torus knots in the universal form \cite{MMM,W1}.
 
 Indeed, consider a torus knot on a three-dimensional sphere, associated with the representation $R$ of a group $G$.  
 The corresponding knot polynomial $P^{[m,n]}_R$, parameterized by two integers $n,m$, is given by the 
 Rosso-Jones formula

\begin{eqnarray}\label{RJ}
P^{[m,n]}_R = \frac{ q^{mn\varkappa_R}}{D_R(q)} \sum_Y \sum_Q
q^{-\frac{n}{m}\varkappa_{_Q}}  \varphi_{_Y} (\bar{\sigma}^{[m,n]}) D_{_Q}(q)
\label{RJ2}
\end{eqnarray}
 where $n$ is the number of strands, 
and $m$ is the "lenght"  (see definitions in e.g. \cite{RJ,MMM}). 
Actually, if $n,m$ are not relatively prime,
we are dealing with a link instead of a knot.
In (\ref{RJ}) $Y$ runs over all Young diagrams with $m$ boxes, and $Q$ runs over all irreducible representations of the gauge group
in one of the subspaces of $R^{\otimes m}$ with the symmetry, associated with the diagram $Y$, with multiplicities taken into account. 
$\varkappa_Q$ is the eigenvalue of the second Casimir on a representation
$Q$, and $D_{_Q}(q)$ denotes the associated quantum dimensions of the same representation.
 We omit definitions of the elements of (\ref{RJ}), which do not depend on the 
gauge group, and hence on the universal parameters, see \cite{RJ,MMM}. 

In \cite{MMM} it is shown, that at least for $[n,m]=[2,m]$ and $[n,m]=[3,m]$, i.e. for two- and three-strand knots/links, the $P^{[m,n]}_R$ can be represented in the universal form,
in case when $R$ is the adjoint representation.  It means that for these cases the group-dependent elements of (\ref{RJ}) - the eigenvalues of the second Casimir operator, as well
as the quantum dimensions, can be written in the universal form. 
For the simplest trefoil knot (i.e. $[n,m]=[2,3])$ the (\ref{RJ}) can be written as:

\begin{multline*}
(uvw)^4\Big(-u^6v^6w^6+(u^6v^6w^5+v^6w^6u^5+w^6u^6v^5)-(u^6v^5w^5+v^6w^5u^5+w^6u^5v^5)- \\
	(u^5v^4w^4+v^5w^4u^4+w^5u^4v^4)
	+(u^5v^4w^3+v^5w^4u^3+w^5u^4v^3+u^5w^4v^3+v^5u^4w^3+w^5v^4u^3)+ \\
	3u^4v^4w^4 -(u^4v^4w^3+v^4w^4u^3+w^4u^4v^3)
	+(u^4v^3w^3+v^4w^3u^3+w^4u^3v^3) -  \\
	(u^4v^2w^2+v^4w^2u^2+w^4u^2v^2)-(u^3v^3w^2+v^3w^3u^2+w^3u^3v^2)+(u^3v^2w^2+v^3w^2u^2+w^3u^2v^2)- \\
	(u^3v^2w+v^3w^2u+w^3u^2v+u^3w^2v+v^3u^2w+w^3v^2u)-2u^2v^2w^2+\\
	(u^2+v^2+w^2)+(uv+vw+wu)+1\Big)
\end{multline*}

\normalsize

where $u=q^\alpha$, $v=q^\beta$ and $w=q^\gamma$, and the connection with (\ref{cad}) is set up by $q=e^{x/2}$.
For the corresponding values of the universal parameters from table \ref{tab:V2}, this expression turns
 into the 
corresponding invariant polynomial - Kauffmann, HOMFLY and exceptional for $so, sl$ and the 
exceptional algebras, respectively 
(see the detailed comparison  in \cite{MMM}). 
To stress the power of the universality one can mention
 that all but one of the coefficients of this universal polynomial is obtained from the requirement that it
  must coincide with the adjoint Kauffmann polynomial for the $SO(N)$ group,
 after which the adjoint HOMFLY 
polynomial for the $SL(N)$ group is automatically recovered!

There are a number (actually series) of representations, besides the adjoint, for which the quantum dimensions
are represented in the universal form. Several applications of these universal 
formulae, such as the representation of (\ref{RJ}) for 
higher-strand knots
 in a universal form, are worth investigating.

In the present paper we study a feature of quantum dimensions, which has been noticed in \cite{AM}, and was later called {\it linear resolvability}  \cite{X2kng}, LR.

 That is, at some points from the Vogel's table
  the universal (quantum) dimensions in (\ref{Z}) and (\ref{X}) may have singularities, thus not yield finite 
  answers. 
 However, the linear resolvability (LR) allows one to give meaning to the universal formula at its singular 
 points. This feature was first noticed and demonstrated in
 \cite{X2kng} in the scope of an investigation of some series of universal quantum dimension formulae.

As the simplest example to demonstrate this phenomenon \cite{AM}, consider the following expression for universal dimension of one of the irreducible representations, appearing
 in the decomposition of the symmetric square of the adjoint representation. It was denoted as $Y_2(\beta)$ by Vogel (see Appendix A), and we refer to its dimension
 similarly as $Y_2(\beta)$:

\begin{multline*}
Y_2(\beta)= \\
- \frac{\left(  2 \alpha-\beta+2\gamma \right) \left( \alpha+2\beta+2\gamma \right)\left( 2\alpha+2\beta+\gamma \right)(\alpha+\beta+\gamma)\left( 3\alpha + 
2\beta+2\gamma \right)\left(2\alpha+2\beta+ 3\gamma \right) }{\beta^2 \alpha \gamma \left( \alpha-\beta \right) \left( \beta - \gamma \right) }
\end{multline*}

At the point, associated to the $\mathfrak {sl}_{2}$ algebra, namely for $\alpha=-2, \beta=2, \gamma=2$, this function is singular. 
Let's check its behavior in the vicinity of this singular point: take $\alpha=-2-y, \beta=2+x, \gamma=2$ for small $y,x$. 
This is the most general deviation, due to the scaling invariance. 
Then, the dimension formula becomes:

\begin{multline*} 
Y_2(\beta)=\\
\frac{(y-2 x-6) (2 y-x) (-y+x+1) (-y+x+2) (-y+x+4) (2 y+x+2)}{x (y+2) (x+2)^2 (y+x+4)}\rightarrow\\
\rightarrow -3\frac{2y-x}{x}
\end{multline*}
for $y\rightarrow 0, x \rightarrow 0$. 

We see that  $Y_2(\beta)$ is singular, since the ratio $(2y-x)/x$ has no limit as one approaches the $(y,x)=(0,0)$ point in an arbitrary way. 
However, when considering this function on the $sl$ line, i.e. when $y=x$, it does have a definite limit 

\begin{eqnarray*}
	Y_2(\beta) \rightarrow -3
\end{eqnarray*}
which is integer and moreover, corresponds to a (virtual) dimension of the adjoint representation $\bold{3}$ of the $\mathfrak {sl}_{2}$ algebra.
 For the quantum dimension in the same limit we get

\begin{eqnarray*}
	-\frac{\sinh(\frac{3x}{2})}{\sinh(\frac{x}{2})}
\end{eqnarray*}
which is the quantum dimension of the corresponding representation with a minus sign. 

So, we see, that the LR feature allows one to make sense out of a universal (quantum) dimension functions even at their
singular points. Moreover, in this example the values obtained in this way are perfectly reasonable and are equal to dimensions of one of the irreducible
 representations of a given algebra (corresponding to
 the point in Vogel's table).

Actually, the situation is even more striking. Some points - most notably, the one, associated with the $\mathfrak {so}_{8}$ algebra, 
with the coordinates $(-2,4,4)$ - belongs to two distinguished lines simultaneously: to the orthogonal and the exceptional lines. 
So, if there is a singularity for some dimension formula, (see below) one may ask whether that singularity is LR, and to which of these lines
we should restrict the function to get a relevant output. 
The answer is - to both! When restricting the function to the orthogonal line, we get an answer, which respects the $Z_2$ group of automorphisms of $D_N$ series of
 Dynkin diagrams, and when restricting it to the exceptional line, we obtain an output, respecting the $S_3$ group of automorphisms of the $\mathfrak {so}_{8}$ algebra. 
 So, in this situation not only we make sense out of the universal function at its singular point, but remarkably obtain two different answers at the same point, each one
  completely relevant!

As an example, consider the same function $Y_2(\beta)$. In the vicinity of the point, corresponding to $\mathfrak {so}_{8}$, i.e. at the $(-2-y,4+2x,4)$ points with small $x,y$,
 $Y_2(\beta)$ tends to 

\begin{eqnarray*}
		Y_2(\beta)\rightarrow 35\frac{x+y}{x}
\end{eqnarray*}

The restriction to the orthogonal line $2\alpha+\beta=0$ corresponds to $x=y$. In this case 
$Y_2(\beta)\rightarrow 70$. 
When restricting the function to the exceptional line $\gamma=2(\alpha+\beta)$, i.e. taking $y=2x$, 
the corresponding dimension is $Y_2(\beta)\rightarrow 105$. 

These outputs can be interpreted as dimensions of representations of $\mathfrak {so}_{8} \rtimes Z_2$ and $\mathfrak {so}_{8} \rtimes S_3$ algebras with 
the Dynkin labels $(0010)\oplus(0001)$ and $(1000)\oplus(0010)\oplus(0001)$, respectively. 
Here we assume the second root to be the central one in the Dynkin diagram, associated to the $\mathfrak {so}_{8}$ algebra; the $S_3$ triality group permutes the other three simple roots,
and the $Z_2$ group of automorphisms acts on the 3-rd and 4-th simple roots.

Thus we demonstrated, that the restriction to each of the appropriate lines leads to a reasonable answer. 
 In both cases one can extend this calculation to the quantum dimension formulae and obtain similar results for the corresponding quantum dimensions. 

Similar calculations can be done for Cartan product of the tensor square of the adjoint and a $Y_2(\beta)$ representations.
 The corresponding universal dimension formula, with permuted coordinates $(\alpha,\beta,\gamma)\rightarrow(\gamma,\beta,\alpha)$
  writes as follows (the general case of arbitrary powers is presented in Section 2):

\begin{multline*}
dim=\\
=\frac{(\alpha +2 \gamma ) (\beta +\gamma ) (2 \beta +\gamma ) (\alpha -\beta -2 \gamma ) (\alpha +\beta +\gamma ) (2 \alpha +\beta +\gamma )}
{\alpha ^3 \beta ^2 \gamma  (\alpha -\beta )^2 }\times \\
\frac{(\alpha +2 \beta +\gamma ) 
(2 \alpha +2 \beta +\gamma ) (2 \alpha -\beta +2 \gamma ) (\alpha +\beta +2 \gamma ) (2 \alpha +\beta +2 \gamma ) (3 \alpha -2 (\beta +\gamma )) (\alpha +2 (\beta +\gamma ))}
{(3 \alpha -\beta ) (\alpha -2 \gamma ) (\alpha -\gamma ) (2 \alpha -\gamma ) (\beta -\gamma )
}
\end{multline*}

It is singular at the $\mathfrak {so}_{8}$ point, and in its vicinity $(4,4+x,-2+y)$ we have 

\begin{eqnarray*}
	dim \rightarrow 35 \frac{y(x-2y)}{x^2}
\end{eqnarray*}

The singular multiplier is equal to -1 on the orthogonal line (when $x=-2y$) and to -3 on the exceptional one (when $x=-y$), and the virtual dimensions appearing
 here are associated with the 
$(2000)$ and $(2000)\oplus(0020)\oplus(0002)$ representations, respectively.

It was proved in \cite{AM} that
the universal formula of quantum dimensions of Cartan products of arbitrary powers of the adjoint and $X_2$ representations\footnotemark  $X(x,k,n)$ \cite{AM,X2kng}
is LR at the points from the Vogel's table.

\footnotetext{ From now on $X(x,k,l)$ denotes any of the $X(x,k,l,\alpha,\beta,\gamma)$ functions with all permutations of $(\alpha,\beta,\gamma)$ parameters. The same holds for $Z(x,k,l)$. }

There is another series of universal quantum dimensions - of Cartan products of arbitrary powers of the adjoint and $Y_2(\beta)$ representations - 
the $Z(x,k,l)$ \cite{{M16QD}}. The universal formulae for dimensions of these series of representations were derived in \cite{LM1}.

The main result of the present paper is the proof of LR for $Z(x,k,l)$ at the points of Vogel's table and at some other distinguished points, classified in \cite{RoadM}.

 In Section 1 we present the expression for the $Z(x,k,l)$ and $X(x,k,n)$ functions.

 Section 2 is devoted to the exact definition of LR and some preliminary notes, essential for the proof of the main statement (Proposition 1).
 
 The outline of the proof is presented in Appendix B, where the most demonstrative examples are chosen.
 We also make a conjecture on the values of functions at their singular points, corresponding to simple Lie algebras,
  when restricting those on the distinguished lines of Vogel's table.

In Section 3 the same statement is proved  for some other distinguished points - $X_1, X_2, E_{7\frac{1}{2}}$ - belonging to the physical region
of the Vogel's plane, i.e. region, where the universal
 dimension of the adjoint representation (\ref{cad}) yields positive outputs (dimensions).

Regarding the remaining set of distinguished points, the so-called $Y$-objects,
 we verify that at exactly three points among them, the $Z(x,k,l)$ and $X(x,k,l)$ are LR, and for two of them these functions
  yield integer valued outputs when 
 considering the $x\rightarrow0$ limit; thus we single out two $Y$-objects  possessing
  additional "algebra-like" quality.

\section{Universal quantum dimension \texorpdfstring{$Z(x,k,l,\alpha,\beta,\gamma)$}{Lg}}

The main object to be examined below for LR at the points from the Vogel's table (and at some other points) is the universal expression of the quantum dimension of Cartan 
products of arbitrary powers of the adjoint and $Y_2(\beta)$ representations \cite{M16QD}. It 
writes as follows\footnotemark :

\begin{multline}\label{Z}
Z(x,k,l,\alpha,\beta,\gamma)= \\
    \sinh\left[\frac{x}{4}: \right. \prod_{i=1}^{k+l} \frac{(\alpha(3-i)+2\gamma)\cdot(\alpha(4-i)+\beta+2\gamma)\cdot(\alpha(3-i)+2\beta+\gamma)}{(\alpha(1-i)+2\beta)\cdot(-\alpha i+\beta)\cdot(\alpha(1-i)+\gamma)} \times \\
    \prod_{i=1}^{k} \frac{\alpha(i-1)-2\beta}{(\alpha i)} \times  \prod_{i=1}^{k+2l} \frac{\alpha(4-i)+2\beta+2\gamma}{\alpha(3-i)+2\gamma} \times \\
    \prod_{i=1}^{l} \frac{(\alpha(3-i)-\beta+2\gamma)\cdot(\alpha(3-i)+\beta+\gamma)\cdot(\alpha(4-i)+2\gamma)}{(\alpha(1-i)-\beta+\gamma)\cdot(\alpha(1-i)+\beta)\cdot(-\alpha i)} \times \\
    \frac{(\alpha(3-2k-2l)+2\beta+2\gamma)\cdot(\alpha(3-2l)+2\gamma)\cdot(\alpha(3-k-2l)+\beta+2\gamma)\cdot(-\alpha k+\beta)}{ \beta \cdot (3\alpha+2\beta+2\gamma)\cdot(3\alpha+2\gamma)
    \cdot(3\alpha+\beta+2\gamma)} 
       \end{multline} 
        \footnotetext{Here the following notation is used:

\begin{eqnarray*}
\sinh\left[x: \right. \,\, \frac{A\cdot B...}{M\cdot N...}\equiv \frac{\sinh(xA)\sinh(xB)...}{\sinh(xM)\sinh(xN)...}
\end{eqnarray*}
 For example

\begin{eqnarray*}
\sinh\left[\frac{x}{4}: \right. \,\, \frac{1\cdot 4}{2}= \frac{\sinh(\frac{x}{4})\sinh(\frac{4x}{4})}{\sinh(\frac{2x}{4})}
\end{eqnarray*}
}

This function, like all other universal formulae, makes sense under permutation of $(\alpha,\beta,\gamma)$ parameters, meaning that it yields
 quantum dimensions of some other representations of the associated simple Lie algebra, which corresponds to the point under consideration with some possible permutation of its 
 coordinates.
 
We also present the formula for universal dimension of Cartan products of arbitrary powers of $X_2$ and the adjoint representation \cite{X2kng}:

\begin{multline}\label{X}
	X(x,k,n,\alpha,\beta,\gamma)=\\ 
	\sinh\left[\frac{x}{4}: \right.\prod _{i=0}^{k-1} \frac{(\alpha  (i-2)-2 \beta )^2(\alpha  (i-2)-2 \gamma )^2(- \alpha (i-2)+\beta +\gamma )^2}{(\alpha  (i+1) )^2
		( -\alpha  (i-1)+\beta)^2(-\alpha  (i-1)+\gamma)^2}\times \\
	\times \prod _{i=0}^{n} \frac{(\alpha  (i+k-2)-2 \beta )(\alpha  (i+k-2)-2 \gamma )(-\alpha  (i+k-2)+\beta +\gamma)}
	{(\alpha (i+k+1))(-\alpha  (i+k-1)+\beta )(-\alpha(i+k-1)+\gamma )} \times \\
	\times \prod _{i=1}^{2k+n} \frac{(\alpha  (i-3)-\beta -2 \gamma )(\alpha  (i-3)-2 \beta -\gamma )(\alpha 
		(i-5)-2 \beta -2\gamma ))}
	{(\alpha  (i-2)-2 \beta )(\alpha  (i-2)-2 \gamma )(-\alpha  (i-2)+\beta +\gamma)} \times \\
	\times \frac{(\alpha +\beta)(\alpha +\gamma)(\alpha  (n+1))}{(2 \alpha +2 \beta )(2 \alpha +2 \gamma )(2 \alpha +\beta
		+\gamma )}\times \\
	\times \frac{(\alpha  (3 k+n-4)-2 \beta -2 \gamma)(\alpha  (3 k+2 n-3)-2
		\beta -2 \gamma )}{(3 \alpha +2 \beta +2 \gamma )(4 \alpha +2 \beta +2
		\gamma )}
\end{multline}

\section{Definition of LR for universal formulae and the main statement}

In this section we give the definition of LR and set out the method, which has been used for the proof of the main result - Proposition 1.

{\bf Definition. }
{\it
 A multivariable function is said to be LR at its singular point, if it yields finite output when approaching that point through all (regular, by definition) but a finite number of (irregular) lines.}

Note, that all known universal (quantum) dimension formulae, particularly (\ref{Z}), are ratios of 
a special form, where both the numerator and denominator decompose into products of a finite number of 
(sines of) linear functions of parameters $(\alpha,\beta,\gamma)$, so that at their singular points some of the factors of the denominator are necessarily zeroing.
 
This feature allows us to prove the following

{\bf Lemma.}
{\it A universal formula is LR at its singular point iff the number of zeroing factors 
 in the denominator is less or equal to those in the numerator at that point. }

 \begin{proof}
Suppose for a universal formula $F$ the number of zeroing terms in the numerator and denominator is $n$ and $d$ respectively.
If $d>n$ then approaching the singular point through lines, other than those given by the equations coinciding with any of the zeroing factors in the numerator, the formula 
obviously yields an infinite output. As the number of such choices is infinite, then $F$ is not LR.

Now suppose $d \leq n$. If we approach the singular points through all but the line given by the equation coinciding with any of the $d$ factors, we will
 necessarily get a finite output for $F$, which means that it is LR. It is clear, that as long as $d<n$, $F$ yields zero, when restricted at any regular line. 
\end{proof}

{\bf Remark 1.}
Obviously, when considered a universal formula on any regular line, both $n$ and $d$ do not change.
 It means that the complete examination of LR can be made by observing the function on a single regular line.
 
{\bf Remark 2.} 
All irregular lines for a given universal formula are exactly determined by each of the factor in its denominator; if there is, 
say a $ c_1\alpha+c_2 \beta+c_3\gamma$ factor in the denominator of the the universal formula, it cannot yield a finite output, when 
restricted on the associated  $c_1\alpha+c_2 \beta+c_3\gamma=0$ line, meaning, that each of the factor of the denominator determines 
an irregular line of the corresponding function. 

{\bf Remark 3.} 
Based on the previous remark, one can easily check, that any of the $sl, so, exc$ lines (see Table 5) is regular for (11) and (12) formulae.

\begin{table}[ht]
	\caption{Isolated solutions in the physical region of Vogel's plane }
	\centering
	\begin{tabular}{|r|r|r|r|r|} \hline
		$\alpha \beta \gamma$ & Dim & Rank & Algebra\\ 
		\hline
		-6	-10	1	&	248	&	8	&	$	E_8	$		\\
		-8	1	-5	&	190	&	8	&	$	E_{7\frac{1}{2}}	$	\\
		-4	1	-7	&	156	&	8	&	$	X_1	$	\\
		-6	-4	1	&	133	&	7	&	$	E_7	$	\\
		1	-3	-5	&	99	&	7	&	$	X_2	$	\\
		-3	-4	1	&	78	&	6	&	$	E_6	$	\\
		-6	2	-5	&	52	&	4	&	$	F_4	$	\\
		3	-5	-4	&	14	&	2	&	$	G_2	$	\\
		\hline
	\end{tabular}  \label{tab:isp}
	\end{table}

\newpage

The {\bf Lemma} is of essential importance for the proof of the following

{\bf Proposition 1.}

{\it  At the points from the Vogel's table the function (\ref{Z}) and functions, obtained from it by all possible permutations of the corresponding 
parameters $(\alpha, \beta, \gamma)$, are LR 
for any set $(k,l)$ with integer non-negative numbers $k,l$. }

Proof is carried out by case by case (for each algebra and for each permutation) examination of the structure of (\ref{Z}), restricting it on the corresponding line
and tracking all possible zero factors appearing both in the numerator and the denominator. In fact, the procedure of the proof automatically highlights all
possible singular points.
Particularly, it turns out that there is an infinite number or series of singular points, (see Appendix B). However, the patterns, governing the
appearance of them is pretty complicated, so we do not find it reasonable to classify them in the scope of this paper.

The main steps, which cover all principal points of the proof, namely, the key cases of the permutations of the parameters for
each of the algebra, are presented in Appendix B.

Finally, we propose a conjecture:

{\bf Conjecture.}

{\it The values of functions $X$ and $Z$, calculated at the singular points by restricting the functions to the corresponding $sl, so, sp$ or 
$exc$ lines,
 are equal to the quantum dimensions of some representations
 of the corresponding algebra.
 Particularly, if a singular point belongs to two distinguished lines simultaneously, the same statement is true for each of the obtained values.}

This conjecture has been checked in a number of cases.

\section{\texorpdfstring{$E_{7\frac{1}{2}},  X_1$}{Lg},  \texorpdfstring{$X_2$}{Lg} and \texorpdfstring{$Y_n$}{Lg} points in Vogel's plane}

Besides the points, corresponding to the simple Lie algebras, there are other distinguished ones in the Vogel's plane - the ones, which along with those from the Vogel's table
 have been obtained in \cite{RoadM} (see also \cite{AM-dubna}, some of them were obtained earlier in \cite{W3,LM2}) from the requirement for the 
universal quantum dimension of the adjoint representation (\ref{cad}) to be regular function of $x$ in the finite complex plane.
 In other words, the quantum dimension, associated to these points, rewrites as 
 a finite sum of exponents. These points are listed in Tables \ref{tab:isp}, \ref{tab2} and \ref{tab3}, along with the points, which correspond to the exceptional simple Lie algebras. 
Note the $E_{7\frac{1}{2}},  X_1$ and  $X_2$ points there, which belong to the physical region of Vogel's plane; they were 
suggested to have the following interpretations:  $E_{7\frac{1}{2}}$, with dimension $190$ and rank $8$, is proven  to be the
 semidirect product of $e_7$ and $H_{56}$ - (56+1)-dimensional Heisenberg algebra \cite{W3, LM2}, 
 $X_1$, with dimension 156 and rank 8, is proposed to be $\mathfrak {so}_{14}\rtimes H_{64}$ semidirect product, and $X_2$ is proposed
 to be the $\mathfrak {so}_{12}\rtimes H_{32}$ semidirect product \cite{W3,LM2,RoadM}.

Examining the behavior of $Z(x,k,l)$ (\ref{Z}) and $X(x,k,l)$ (\ref{X})  \cite{X2kng} functions at these points we present the following

{\bf Proposition 2.}
{\it At the $X_1, X_2,E_{7\frac{1}{2}}$ points in the Vogel's plane, both $Z(x,k,l)$ and $X(x,k,l)$ functions are LR.}

Proof is straightforward.

The remaining 48 points, corresponding to the so called Y-objects, are given in Tables \ref{tab2} and 
 \ref{tab3}. Dimensions of their "adjoint representation" ,
 i.e. values of $f(x)$ at the associated points, when $x \rightarrow 0$, are negative\footnote{Note some irregularity in the notations: 
there is an object  $Y_6^{\prime}$, which stands out from the remaining ones ($Y_i, i=1,2,...,47$) in its notation. The reason is that in \cite{RoadM}  two different solutions of 
Diophantine equations were accidentally denoted by the same notation $Y_6$, and here, trying to have minimal changes in notations, we denote one of them as $Y_6^{\prime}$.}.

We tested both $Z(x,k,l)$ (\ref{Z}) and $X(x,k,l)$ (\ref{X}) \cite{X2kng} formulae on LR at those points and got the following result:

{\bf Proposition 3.}
{\it At the points $Y_2, Y_6, Y_{32}$ from Table \ref{tab2} both $Z(x,k,l)$ and $X(x,k,l)$ formulae 
are regular. At all other points from the same table it is  possible to choose a $(k,l)$ pair, for which either $Z(x,k,l)$ or $X(x,k,l)$ is singular and 
not LR for some permutation of the Vogel's parameters.}

{\bf Proof.}
The desired result follows from direct substitution of the corresponding sets of parameters $(\alpha,\beta,\gamma)$ (with all possible permutations) into the denominators of 
 $Z(x,k,l)$ and $X(x,k,l)$.

{\bf Remark.}
A notable fact is that when taking the $x\rightarrow 0$ limit, both
 $Z(x,k,l)$ and $X(x,k,l)$ formulae yield integer valued outputs at the 
$Y_2$ and $Y_{32}$ points, pointing out a remarkable similarity of those "unknown" objects with
the simple Lie algebras.

As an example of appearance of a singularity let us take the $Y_3$ point, with $(\alpha,\beta,\gamma)=(6,4,5)$ and
consider the $Z(x,k,l,\beta,\gamma,\alpha)=Z(x,4,1,4,5,6)$ function at that point.
Evidently, when $k+2l \geq 6$, the $6$-th multiplier of
$$ \sinh\left[\frac{x}{4}: \right.\prod_{i=1}^{k+2l} \frac{\alpha(4-i)+2\beta+2\gamma}{\alpha(3-i)+2\gamma}$$ 
produces a zero in the denominator: $\alpha(-3)+2\gamma\rightarrow 4(-3)+2 \times 6=0$, while 
for the same choice of $(k,l)$ there is not a multiplier in the numerator to cancel it out, which means that $Z(x,4,1,\beta,\gamma,\alpha)$ is singular at the $Y_{3}$ point.

\section*{Conclusion}
The ultimate result of this work is the proof of the statement that at all points, corresponding to the classical and the exceptional algebras, and also for three additional distinguished points in 
the physical region of the Vogel's plane, all known universal formulae for quantum dimensions are LR.
This means that at all these points universal formulae yield finite and relevant answers either directly or after approaching the singular point through a corresponding 
classical or exceptional line if needed. Note, that for
algebras which lie on the intersection of such lines, such as $\mathfrak {so}_{8}$, both lines make sense, so that in such cases we get two relevant outputs from a single function at a given singular point. 

Another non-trivial statement is that among so-called Y-objects, with corresponding points belonging to the non-physical region of the 
Vogel's plane, there are another three points, at which all universal 
quantum dimensions are regular. Furthermore,
we observe that two of them behave like real existing algebras, in the sense that
the universal dimension formulae yield integer-valued output at those points.

 There is an additional universal formula for dimension (not the quantum dimension), given by 6.10 in \cite{LM1}, which may be studied on LR.
 There is not a quantum dimension generalization of it, and it seems to show an irregular behavior, meaning that it is singular even without  permutation of the
 parameters, which is not the case for any other universal formula. It needs
to be checked, as well as its quantum analogue is yet to be derived and checked on LR.

Present results give rise to a number of natural questions, such as what is the underlying
reason of universal quantum dimensions possessing the LR feature, as well as where does the remarkable
 property of $Y_2$ and $Y_{32}$ points 
inducing integer-valued outputs from universal 
dimension formulae come from.
Note, that none of these $Y$-objects coincide with those, connected with equivelar maps on genus two surfaces, see \cite{KhM16}.

\begin{table}[ht]
\caption{Isolated solutions in the non-physical region of Vogel's plane} 
      \centering
		\begin{tabular}{|c|c|c|c|} \hline
$\alpha \beta \gamma$ & Dim & Rank & Notation\\ \hline

1	1	1	&	-125	&	-19	&	$	Y_1	$      \\
10	8	7	&	-129	&	-1	&	$	Y_2	$	\\
6	4	5	&	-130	&	-4	&	$	Y_3	$	\\
2	2	3	&	-132	&	-10	&	$	Y_4	$	\\
5	7	8	&	-132	&	-2	&	$	Y_5	$	\\
5	8	6	&	-132	&	-2	&	$	Y_6	$	\\
4      5       3      &       -133&       -2     &      $      Y_6^{'} $       \\
4	7	5	&	 -135&	-3	&	$	Y_7	$       \\
7	6	4	&	-135	&	-3	&	$	Y_8	$	\\
2	4	3	&	-140	&	-8	&	$	Y_9	$	\\
2	1	2	&	-144	&	-14	&	$	Y_{10}$	\\
2	1	1	&	-147	&	-17	&	$	Y_{11}	$	\\
7	3	4	&	-150	&	-4	&	$	Y_{12}	$	\\
2	4	5	&	-153	&	-7	&	$	Y_{13}	$	\\
5	3	2	&	-153	&	-7	&	$	Y_{14}	$	\\
1	2	3	&	-165	&	-13	&	$	Y_{15}	$	\\
2	6	5	&	-168	&	-6	&	$	Y_{16}	$	\\
6	2	7	&	-184	&	-6	&	$	Y_{17}	$	\\
4	5	13	&	-186	&	-2	&	$	Y_{18}	$	\\
3	10	4	&	-186	&	-4	&	$	Y_{19}	$	\\
3	7	2	&	-187	&	-7	&	$	Y_{20}	$	\\
1	1	3	&	-189	&	-17	&	$	Y_{21}	$	\\
11	5	3	&	-189	&	-3	&	$	Y_{22}	$	\\
4	1	3	&	-195	&	-11	&	$	Y_{23}	$	\\
2	1	4	&	-195	&	-13	&	$	Y_{24}	$	\\
3	11	4	&	-200	&	-4	&	$	Y_{25}	$	\\
2	3	8	&	-207	&	-7	&	$	Y_{26}	$	\\
2	5	9	&	-207	&	-5	&	$	Y_{27}	$	\\
3	1	5	&	-221	&	-11	&	$	Y_{28}	$	\\ 
\hline

	\end{tabular} \label{tab2}
	 \end{table}

\newpage

\begin{table}[ht]\caption{Table 2 continued}
\centering
		\begin{tabular}{|c|c|c|c|c|} \hline
$\alpha \beta \gamma$ & Dim & Rank & Notation\\ \hline

1	4	5	&	-228	&	-10	&	$	Y_{29}	$	\\
2	1	5	&	-231	&	-13	&	$	Y_{30}	$	\\
4	1	1	&	-242	&	-18	&	$	Y_{31}	$	\\
6	5	22	&	-244	&	-2	&	$	Y_{32}	$	\\
18	4	5	&	-245	&	-3	&	$	Y_{33}	$	\\
14	4	3	&	-247	&	-5	&	$	Y_{34}	$	\\
10	2	3	&	-252	&	-8	&	$	Y_{35}	$	\\
1	4	6	&	-252	&	-10	&	$	Y_{36}	$	\\
3	5	16	&	-258	&	-4	&	$	Y_{37}	$	\\
6	1	2	&	-272	&	-14	&	$	Y_{38}	$	\\
1	3	7	&	-285	&	-11	&	$	Y_{39}	$	\\
1	5	7	&	-285	&	-9	&	$	Y_{40}	$	\\
14	2	5	&	-296	&	-6	&	$	Y_{41}	$	\\
6	8	1	&	-319	&	-9	&	$	Y_{42}	$	\\
1	3	8	&	-322	&	-12	&	$	Y_{43}	$	\\
4	1	9	&	-342	&	-10	&	$	Y_{44}	$	\\
10	1	4	&	-377	&	-11	&	$	Y_{45}	$	\\
12	1	5	&	-434	&	-10	&	$	Y_{46}	$	\\
1	6	14	&	-492	&	-10	&	$	Y_{47}	$       \\ 
\hline
\end{tabular}   \label{tab3}
\end{table}

\section*{Acknowledgments}

The work of MA was fulfilled within the Regional Doctoral Program on Theoretical and Experimental Particle Physics 
sponsored by VolkswagenStiftung.
The work of MA and RM is partially supported by the Science Committee of the Ministry of Science 
and Education of the Republic of Armenia under contract  20AA-1C008.

\section{Appendix A. Universal description of simple Lie algebras}
The Vogel's parameterization of simple Lie algebras is presented in Table \ref{tab:V2}.

\begin{table}[ht]
	\caption{Vogel's parameters and distinguished lines}
	\centering
	\begin{tabular}{|r|r|r|r|r|r|} 
		\hline Algebra/Parameters & $\alpha$ &$\beta$  &$\gamma$  & $t$ & Line \\ 
		\hline  $\mathfrak{sl}_N$  & -2 & 2 & $N$ & $N$ & $\alpha+\beta=0, sl$ \\ 
		\hline $\mathfrak{so}_N$ & -2  & 4 & $N-4$ & $N-2$ & $ 2\alpha+\beta=0, so$ \\ 
		\hline  $ \mathfrak{sp}_N$ & -2  & 1 & $N/2+2$ & $N/2+1$ & $ \alpha +2\beta=0,sp$ \\ 
		\hline $exc(n)$ & $-2$ & $2n+4$  & $n+4$ & $3n+6$ & $\gamma=2(\alpha+\beta), exc$\\ 
		\hline 
	\end{tabular}
	
	{For the exceptional 
		line $n=-2/3,0,1,2,4,8$ for $\mathfrak {g}_{2}, \mathfrak {so}_{8}, \mathfrak{f}_{4}, \mathfrak{e}_{6}, \mathfrak {e}_{7},\mathfrak {e}_{8} $, 
		respectively.} \label{tab:V2}
\end{table}

To give an idea on the origin of this table we write the following universal (i.e. valid for any simple Lie algebra) decomposition of the symmetric square of the adjoint representation \cite{V0}:

\begin{eqnarray}\label{sad}
S^2 \mathfrak{g}=1 \oplus Y_2(\alpha) \oplus Y_2(\beta) \oplus Y_2(\gamma)
\end{eqnarray}

Let $2t$ denote the eigenvalue of the second Casimir operator on the adjoint representation $\mathfrak{g}$, then the eigenvalues of the same operator on representations in 
(\ref{sad}) we denote as  $4t-2\alpha, 4t-2\beta, 4t-2\gamma$, correspondingly. In this way we define $\alpha, \beta, \gamma$ (Vogel's) parameters. It can be proved \cite{V0}  that with these definitions  $\alpha+\beta+\gamma=t$.

\section{Appendix B. Proof of LR of \texorpdfstring{$Z(x,k,l)$}{Lg} formula}

The procedure of the proof is carried out in the following way: first, we take the main formula (\ref{Z}), and for each of the point $(\alpha, \beta,\gamma)$ from Table \ref{tab:V2}
in the Vogel's plane, including those obtained by another 5 permutations of the parameters, 
 examine its expression in the vicinity of the point in question, restricting it on the corresponding distinguished line (Table \ref{tab:V2}) beforehand.
 Then, we trace the number of zeroing factors in both its numerator ($n$) and denominator ($d$) at the corresponding points.
 Based on the Lemma (Section 1), the proof of LR is in fact equivalent to the checking of
  the realization of the $n\geq d$ inequality in each of the possible cases, namely,
 for every possible non-negative integer valued set $(k,l)$ for each of the permutations of the corresponding Vogel's parameters.
 
 Since the implementation of this procedure is quite repetitive, we find it reasonable to
 present the explicit calculations for several key cases only, which are sufficient to outline the essence of the proof.
 They are presented in the following section.

\paragraph{Classical algebras.}

\subsection{\texorpdfstring{$A_N$}{Lg}}
\subsubsection{\texorpdfstring{$\alpha,\gamma,\beta$}{Lg}}
Here we examine the $Z(x,k,l,\alpha,\gamma,\beta)$
for the parameters, corresponding to the $A_N$ algebra, by presenting the corresponding formulae, which are obtained by
every possible choice of the set $k,l$:

\subsubsection{\texorpdfstring{$l=0,k=1$}{Lg}}
\begin{equation*}
Z(x,1,0,-2,N+1,2)=\sinh\left[\frac{x}{4}: \right.\frac{2N+4}{2^2}
\end{equation*}

\subsubsection{\texorpdfstring{$l=0,k>1$}{Lg}}
\begin{equation*}
Z(x,k,0,-2,N+1,2)=\sinh\left[\frac{x}{4}: \right.\frac{(2N)\cdot(2N+4k)}{(2k)^2}\times \frac{(2N+2)^2\cdot(2N+4)^2\dots(2N+2k-2)^2}
{2^2\cdot4^2\dots (2k-2)^2}
\end{equation*}

\subsubsection{\texorpdfstring{$l=1,k=0$}{Lg}}
\begin{equation*}
Z(x,0,1,-2,N+1,2)=\sinh\left[\frac{x}{4}: \right.\frac{(2N)\cdot(2N+4)}{2^2}
\end{equation*}

\subsubsection{\texorpdfstring{$l=1,k\geq1$}{Lg}}
\begin{equation*}
Z(x,k,1,-2,N+1,2)=\sinh\left[\frac{x}{4}: \right.\frac{(2N)\cdot(2N+4k+4)}{(2k+2)^2}\times \frac{(2N+2)^2\cdot(2N+4)^2\dots(2N+2k)^2}
{2^2\cdot4^2\dots (2k)^2}
\end{equation*}
Obviously, each of the functions written above is regular for any $N\in \mathbb{N}$.

Let's move on to the remaining $(k,l)$ sets:

\subsubsection{\texorpdfstring{$l=2,k=0$}{Lg}}
\begin{multline*}
Z(x,0,2,-2,N+1,2)=\\
\sinh\left[\frac{x}{4}: \right.
\frac{(2\alpha+2\beta)\cdot(N-1)\cdot N \cdot (N+1)^2\cdot (N+7) \cdot(2 N+2)\cdot (2 N+6)\cdot (2 N+8)}{2^2\cdot4^3\cdot(N-3)\cdot (N+3)^2\cdot (N+5)}
\end{multline*}
We see, that there is a $2\alpha+2\beta$ factor, which is zeroing at any point of the 
$2\alpha+2\beta=0$ line, so that one can easily determine, that the possible number of zeroing factors in the numerator is
always greater or equal to those in the denominator, namely $d\leq n$, which means, that the initial function is LR.

\subsubsection{\texorpdfstring{$l=2,k\geq1$}{Lg}}
\begin{multline*}
Z(x,k,2,-2,N+1,2)=\\
\sinh\left[\frac{x}{4}: \right.\frac{(2\alpha+2\beta)\cdot6\cdot(2N)\cdot(N+1)\cdot(N-1)\cdot(2N+4k+8)}
{2\cdot4\cdot(2k+4)\cdot(N+2k+3)\cdot(N+2k+5)\cdot(N+3)}\times \\ 
\frac{(N+2k+7)\cdot(N+2k+1)\cdot(2N+2k+2)\cdot(2N+2k+6)}{(N-3)\cdot(2k+2)\cdot(2k+4)\cdot(2k+6)}\times\\
\frac{(2N+2)^2\cdot(2N+4)^2\dots(2N+2k)^2}{2^2\cdot\ 4^2\dots (2k)^2}
\end{multline*}

Proof of the LR of this function is similar to that for the previous one.
Notice, that for each of the integer $k\geq1$, the corresponding function has a singularity (linear resolvable, of course), when $N=3$.
This particular case is interesting in the sense, that it explicitly demonstrates, that the set of singularities of the function (\ref{Z}) is basically infinite.

\subsubsection{\texorpdfstring{$l\geq3,k=0$}{Lg}}
\begin{multline*}
Z(x,0,l,-2,N+1,2)=\\
\sinh\left[\frac{x}{4}: \right.\frac{(2\alpha+2\beta)\cdot(2N)\cdot(N+1)^2\cdot(N-1)\cdot(2N+4l)}
{(2l-2)\cdot(2l)^2}\times \\
\frac{(N+4l-1)\cdot(4l-2)}{(N-2l+1)\cdot(2N+2l)\cdot(N+2l-1)^2\cdot(N+2l+1)} \times \\
\frac{(2N+2)\dots(2N+4l-2)}{2\cdot\ 4\dots (4l-2)}
\end{multline*}

\subsubsection{\texorpdfstring{$l\geq3,k\geq1$}{Lg}}
\begin{multline*}
Z(x,0,3,-2,N+1,2)=\\
\sinh\left[\frac{x}{4}: \right.\frac{(2\alpha+2\beta)\cdot(N+1)\cdot(N-1)\cdot(2N+4k+4l)}
{(2l-2)\cdot(2l)\cdot(2k+2l)\cdot(N+2l-1)}\times \\
\frac{(N+2k+4l-1)\cdot(N+2k+1)\cdot(4l-2)}{(N-2l+1)\cdot(2N+2k+2l)\cdot(N+2k+2l-1)\cdot(N+2k+2l+1)} \times \\
\frac{(2N+2)\cdot(2N+4)\dots(2N+2k)}{2\cdot\ 4\dots 2k} \times 
\frac{2N\cdot(2N+2)\dots(2N+2k+4l-2)}{2\cdot\ 4\dots (2k+4l-2)}
\end{multline*}

The same reasoning, which proves the LR, holds for the latter two cases.

Thus, we proved the LR of the $Z(x,k,l,\alpha,\gamma,\beta)$ function at any $N \in \mathbb{N}$ point lying on the $sl$ line.

\subsection{\texorpdfstring{$B_N$}{Lg}}
\subsubsection{\texorpdfstring{$\beta,\gamma,\alpha$}{Lg}}
Let's prove, that the $Z(x,k,l,4,2N-3,-2)$ function is LR for any non-negative integer set $(k,l)$.
To prove the LR of $Z(x,k,l,\beta,\gamma,\alpha)$  at the $(4,2N-3,-2)$ points, where $N \in \mathbb{N}$, we examine it on the $so$ line in the following cases:
\subsubsection{\texorpdfstring{$l=1,k=0$}{Lg}}
In this case $Z$ writes as follows
\begin{equation*}
Z(x,0,1,4,2N-3,-2)=\sinh\left[\frac{x}{4}: \right.\frac{(4N)(4N-2)(2N+3)}{2\cdot 4\cdot (2N-1)}
\end{equation*}
it is regular on the $so$ line for any integer $N$.

One can easily determine, that following 4 functions are also regular for any integer $N$.

\subsubsection{\texorpdfstring{$l=1,k\geq1$}{Lg}}
\begin{multline}
Z(x,k,1,4,2N-3,-2)=\\
=
\sinh\left[\frac{x}{4}: \right.
\frac{(2\alpha+\beta)\cdot4\dots (4k-4)}{6\cdot 10\cdot 14\dots (4k+2)} \times \\
\frac{(2N+5)\cdot(2N+1)\dots (2N-4k+5)}{(2N-7)\cdot(2N-11)\dots (2N-4k-7)} \times 
\frac{4N\cdot(4N-4)\dots (4N-4k)}{(4N-6)\cdot (4N-10)\dots (4N-4k-6)}
\times \\
\frac{(4N-6)^2\cdot(4N-10)^2\dots(4N-4k-2)^2}{4^2\cdot8^2\cdot \dots (4k)^2}\times \\
\frac{(4N-2)\cdot(2N-4k-3)^2\cdot(4N-8k-6)\cdot(2N-7)\cdot(2N+3)}{2\cdot4\cdot(2N-3)^2\cdot(2N+5)\cdot(1-2N)} .
\end{multline}

\subsubsection{\texorpdfstring{$l=2,k=0$}{Lg}}
\begin{equation*}
Z(x,0,2,4,2N-3,-2)=\\
=
-\sinh\left[\frac{x}{4}: \right. \frac{(4N)\cdot(4N-4)\cdot(4N-2)\cdot(2N+1)\cdot(4N-14)}{2\cdot4\cdot6\cdot8\cdot(2N-7)}
\end{equation*}

\subsubsection{\texorpdfstring{$l=2,k\geq1$}{Lg}}
\begin{multline}
Z(x,k,2,4,2N-3,-2)=\\
=
-\sinh\left[\frac{x}{4}: \right.
\frac{4\dots (4k)}{10\cdot 14\dots (4k+6)} \times \\
\frac{(2N+5)\cdot(2N+1)\dots (2N-4k+1)}{(2N-7)\cdot (2N-11)\dots (2N-4k-11)} \times 
\frac{4N\cdot(4N-4)\dots (4N-4k-4)}{(4N-6)\cdot (4N-10)\dots (4N-4k-10)}
\times \\
\frac{(4N-6)\cdot(4N-10)\dots(4N-4k-2)}{4\cdot8\cdot \dots 4k}\times \\
\frac{(4N+2)\cdot (4N-2)\cdot\dots(4N-4k-10)}{2\cdot 6\cdot 4\cdot 8  \dots (4k+8)}\times \\
\frac{(2N-4k-3)\cdot(2N-4k-11)\cdot(4N-8k-14)}{(2N-3)\cdot(2N+5)\cdot (4N+2)} .
\end{multline}

\subsubsection{\texorpdfstring{$l\geq3,k=0$}{Lg}}

\begin{multline}
Z(x,0,l,4,2N-3,-2)=\\
=
\sinh\left[\frac{x}{4}: \right.
\frac{(4N)\cdot(4N-4)\dots(4N+4-4l)}{2\cdot6 \cdot \dots(4l-2)}\times \\
\frac{(2N+5)\cdot (2N+1)\cdot\dots(2N+9-4l)}{(2N-7)\cdot (2N-11) \dots (2N-4l-3)}\times \\
\frac{(4N-2-4(l+1))\dots(4N-2-4(2l-2))}{(4(l-1))\cdot(4l)\dots (4(2l-2))} \times \\ 
\frac{4\cdot (2\alpha+\beta)}{(4l-8)\cdot(4l-4)\cdot 4l} \times \frac{(2N+1)\cdot(2N+5)\dots (2N+4l-11)}{(2N+7)\cdot(2N+11)\dots (2N+4l-5)} \times  \\ 
\frac{(2N-5)\cdot(2N-9)\dots (2N+7-4l)}{(2N-11)\cdot(2N-15)\dots (2N+1-4l)} \times \\
\frac{(4N-2)\cdot(2N-4k-3)\cdot(2N-8l+5)\cdot (8l-8) \cdot 
	(4N-8l+2)}{(2N-3)\cdot(2N+5)}.
\end{multline}

\subsubsection{\texorpdfstring{$l\geq3,k\geq1$}{lg}}

\begin{multline}
Z(x,k,l,4,2N-3,-2)=\\
=
\sinh\left[\frac{x}{4}: \right.
\frac{(4N)\cdot(4N-4)\dots(4N+4-4k-4l)}{2\cdot6 \cdot \dots(4k+4l-2)}\times \\
\frac{(2N+5)\cdot (2N+1)\cdot\dots(2N+9-4k-4l)}{(2N-7)\cdot (2N-11) \dots (2N-4k-4l-3)}\times \\
\frac{(4N-2-4(k+l+1))\dots(4N-2-4(k+2l-2))}{4\cdot 8\dots 4k} \times \frac{4\cdot (2\alpha+\beta)}{(4l-8)\cdot(4l-4)\cdot 4l}
\times \\ \frac{(2N+1)\cdot(2N+5)\dots (2N+4l-11)}
{(2N+7)\cdot(2N+11)\dots (2N+4l-5)} \times  \frac{(4N+2)\cdot(4N-2)\dots (4N-2-4k)}
{(4(k+l-1))\cdot(4(k+l))\dots (4(k+2l-2))} \times \\
\frac{(2N-5)\cdot(2N-9)\dots (2N+7-4l)}{(2N-11)\cdot(2N-15)\dots (2N+1-4l)} 
\times \\ \frac{(2N-4k-3)\cdot(2N-4k-8l+5)\cdot (8l-8) \cdot 
	(4N-8k-8l+2)}{(2N-3)\cdot(2N+5)\cdot(4N+2)}.
\end{multline}

\subsection{\texorpdfstring{$C_N$}{Lg}}
\subsubsection{\texorpdfstring{$\beta,\alpha,\gamma$}{Lg}}
Let's examine the following functions:

\subsubsection{\texorpdfstring{$l=0,k\geq1$}{Lg}}
\begin{multline}
Z(x,k,0,1,-2,N+2)=\\
=
\sinh\left[\frac{x}{4}: \right.
\frac{(2N+5)\cdot(2N+4)\dots(2N+6-k)}{3\cdot4 \cdot \dots(k+2)}\times \\
\frac{(2N+6)\cdot (2N+5)\cdot\dots(2N+7-k)}{4\cdot 5 \dots (k+3)}\times \frac{(k+1)\cdot(k+2)^2\cdot(k+3)}
{1\cdot2^2\cdot 3} \times \\ \frac{(N-k+2)\cdot(N-k+1)\cdot(2N+6-k)\cdot(2N+5-k)\cdot(2N+4-k)}
{(2N+3)\cdot(2N+4)\cdot(2N+5)^2\cdot (2N+6)\cdot(2N+7)}  \times \\
\frac{(2N+5-k)\cdot(2N+7)\cdot(2N+3-2k)}{(N+1)\cdot(N+2)} 
\end{multline}

\subsubsection{\texorpdfstring{$l\geq1,k=0$}{Lg}}
\begin{multline}
Z(x,0,l,1,-2,N+2)=\\
=
\sinh\left[\frac{x}{4}: \right.
\frac{(2N+5)\cdot(2N+4)\dots(2N+6-l)}{3\cdot4 \cdot \dots(l+2)}\times \\
\frac{(2N+6)\cdot (2N+5)\cdot\dots(2N+7-l)}{4\cdot 5 \dots (l+3)}\times \\
\frac{(2N+7)\dots(2N+8-l)}{1\cdot 2\dots l} \times \frac{(2N+8)\cdot(2N+7)\dots(2N+9-l)}{2\cdot3\cdot (l+1)}
\times \\ \frac{(N-l+2)\cdot(N-l+1)\cdot(2N+6-2l)\cdot(2N+5-2l)\cdot(2N+4-2l)}
{(2N+3)\cdot(2N+4)\cdot(2N+5)^2\cdot (2N+6)\cdot(2N+7)} \times \\
\frac{(N+4-l)\cdot(N+3-l)\cdot(2N+5-2l)\cdot(2N+7-2l)\cdot(2N+3-2l)}{(N+1)\cdot(N+2)\cdot(N+3)\cdot(N+4)} 
\end{multline}

\subsubsection{\texorpdfstring{$l\geq1,k\geq1$}{Lg}}
The $Z(x,k,l,1,-2,N+2)$ rewrites as follows:
\begin{multline}
Z(x,k,l,1,-2,N+2)=\\
=
\sinh\left[\frac{x}{4}: \right.
\frac{(2N+5)\cdot(2N+4)\dots(2N+6-k-l)}{3\cdot4 \cdot \dots(k+l+2)}\times \\
\frac{(2N+6)\cdot (2N+5)\cdot\dots(2N+7-k-l)}{4\cdot 5 \dots (k+l+3)}\times \\
\frac{(2N+7)\dots(2N+8-l)}{1\cdot 2\dots l} \times \frac{(2N+8)\cdot(2N+7)\dots(2N+9-l)}{2\cdot3\cdot (l+1)}
\times \frac{(k+1)\cdot(k+2)^2\cdot(k+3)}
{1\cdot2^2\cdot 3} \times \\ \frac{(N-k-l+2)\cdot(N-k-l+1)\cdot(2N+6-k-2l)\cdot(2N+5-k-2l)\cdot(2N+4-k-2l)}
{(2N+3)\cdot(2N+4)\cdot(2N+5)^2\cdot (2N+6)\cdot(2N+7)} \times \\
\frac{(N+4-l)\cdot(N+3-l)\cdot(2N+5-k-2l)\cdot(2N+7-2l)\cdot(2N+3-2k-2l)}{(N+1)\cdot(N+2)\cdot(N+3)\cdot(N+4)} 
\end{multline}

As we see, in all three cases the $Z(x,k,l,1,-2,N+2)$ function is regular for any integer valued set $(k,l)$ and $N \in \mathbb{N}$.

\subsection{\texorpdfstring{$D_N$}{Lg}}
\subsection{\texorpdfstring{$\gamma,\beta,\alpha$}{Lg}}
\subsubsection{\texorpdfstring{$l+k>4$}{Lg} and \texorpdfstring{$l,k\geq1$}{Lg}}
In the following expression $T=2N-4$, $T\geq 4$, since $N\geq4$:
\begin{multline}
Z(x,k,l,T,4,-2)=\\
=
\sinh\left[\frac{x}{4}: \right.
\frac{(4+T)}{(4-3T)\cdot(4-4T)} \times \frac{(4+2T)\cdot(4+3T)\dots(4+(k+l-3)T)}{(4-5T)\cdot(4-6T)\dots(4-(k+l)T)} \times \\
\frac{(2T)\cdot T \cdot (2\alpha+\beta)}{8\cdot(8-T)\cdot(8-2T)\cdot(8-3T)} \times \frac{T\cdot(2T)\dots((k+l-4)T)}{(8-4T)\cdot(8-5T)\dots(8-(k+l-1)T)} \times \\
\frac{(6+2T)\cdot(6+T)\cdot6\cdot(6-T)}{2\cdot(2+T)\cdot(2+2T)\cdot(2+3T)} \times \frac{(6-2T)\cdot(6-3T)\dots(6-(k+l-3)T)}{(2+4T)\cdot(2+5T)\dots(2+(k+l-1)T)} \times \\
\frac{8\cdot(8-T)\dots(8-(k-1)T)}{T\cdot(2T)\dots(kT)} \times \frac{1}{4-2T}\times \\
\frac{(4-2T)\cdot(4-3T)\cdot(4-4T)\dots(4-(k+2l-4)T)}
{(4+3T)\cdot(4+4T)\cdot(4+5T)\dots(4+(k+2l-3)T)} \times \\
\frac{(8-2T)\cdot(8-T)\cdot8\dots(8+(l-3)T)}{4\cdot(4-T)\cdot(4-2T)\dots (4-(l-1)T)}\times \\ \frac{(2+2T)\cdot(2+T)\cdot2\dots(2-(l-3)T)}{6\cdot(6+T)\cdot(6+2T)\dots
	(6+(l-1)T)} \times \\
\frac{(4-3T)\cdot(4-2T)\dots(4+(l-4)T)}{T\cdot(2T)\dots(lT)}\times \\ \frac{(4+(3-2k-2l)T)\cdot(4+(2l-3)T)\cdot((2l+k-3)T)\cdot(kT-4)}{(4-3T)}
\end{multline}

A careful inspection of the above written formula shows, that for any integer valued $T\geq 4$ the number of zeroing factors
in the denominator is not greater than those in the numerator: $d\leq n$, which proves the LR of it.

In the following cases proofs are either evident or repeat those for the previous case.

\subsubsection{\texorpdfstring{$l+k=4$}{Lg} and \texorpdfstring{$k, l\geq1$}{Lg}}
In this case we have the following function:
\begin{multline}
Z(x,k,l,T,4,-2)=\\
=
\sinh\left[\frac{x}{4}: \right.
\frac{4\cdot(4+T)}{(4-3T)\cdot(4-4T)}\times \frac{(3T)\cdot(2T)\cdot T \cdot (2\alpha+\beta)}{8\cdot(8-T)\cdot(8-2T)\cdot(8-3T)} \times \\
\frac{(6+2T)\cdot(6+T)\cdot6\cdot(6-T)}{2\cdot(2+T)\cdot(2+2T)\cdot(2+3T)} \times \frac{8\cdot(8-T)\dots(8-(3-l)T)}{T\cdot(2T)\dots((4-l)T)} \times \\
\frac{(4+3T)\cdot(4+2T)\dots(4-lT)}{(4-2T)\cdot(4-T)\dots(4+(l+1)T)} \times  \frac{(8-2T)\cdot(8-T)\cdot8\dots(8+(l-3)T)}{4\cdot(4-T)\cdot(4-2T)\dots (4-(l-1)T)}\times \\ 
\frac{(2+2T)\cdot(2+T)\cdot2\dots(2-(l-3)T)}{6\cdot(6+T)\cdot(6+2T)\dots(6+(l-1)T)} \times \\
\frac{(4-3T)\cdot(4-2T)\dots(4+(l-4)T)}{T\cdot(2T)\dots(lT)}\times \\
\frac{(4-5T)\cdot(4+(2l-3)T)\cdot((l+1)T)\cdot(4-(4-l)T)}{4\cdot(3T)\cdot(3T-4)\cdot(3T+4)}
\end{multline}
\subsubsection{\texorpdfstring{$l+k=4$}{Lg} and \texorpdfstring{$k=0$}{Lg}}
\begin{multline}
Z(x,0,4,T,4,-2)=\\=
\frac{(2\alpha+\beta)\cdot4\cdot (5T)\cdot(2-T)\cdot(6-T)\cdot(8+T)\cdot(4-5T)\cdot(4+T)}{(3T)\cdot(4T)\cdot(8-3T)\cdot(2+3T)\cdot(4+4T)\cdot(4-3T)\cdot(4+3T)
	\cdot(6+3T)}
\end{multline}

\subsubsection{\texorpdfstring{$l+k=4$}{Lg} and \texorpdfstring{$l=0$}{Lg}}
\begin{multline}
Z(x,4,0,T,4,-2)=\\
=
-\sinh\left[\frac{x}{4}: \right.
\frac{(2\alpha+\beta)\cdot6\cdot (T)\cdot(4+T)\cdot(6+2T)\cdot(6+T)\cdot(6-T)\cdot(4+2T)\cdot(4-5T)}{2\cdot(3T)\cdot(4T)\cdot(4-3T)\cdot(2+T)\cdot(2+2T)\cdot(2+3T)\cdot(4-2T)\cdot(4-T)}
\end{multline}

\subsubsection{\texorpdfstring{$l+k=3$}{Lg} and \texorpdfstring{$l\geq1$}{Lg}}
\begin{multline}
Z(x,k,l,T,4,-2)=\\
=
\sinh\left[\frac{x}{4}: \right.
\frac{(2T)\cdot T}{8\cdot(8-T)\cdot(8-2T)} \times \\
\frac{(6+2T)\cdot(6+T)\cdot6}{2\cdot(2+T)\cdot(2+2T)} \times \frac{8\cdot(8-T)\dots(8-(2-l)T)}{T\cdot(2T)\dots((3-l)T)} \times \\
\frac{(4+3T)\cdot(4+2T)\dots(4-(l-1)T)}{(4-2T)\cdot(4-T)\dots(4+lT)} \times  \frac{(8-2T)\cdot(8-T)\cdot8\dots(8+(l-3)T)}{4\cdot(4-T)\cdot(4-2T)\dots (4-(l-1)T)}\times \\ 
\frac{(2+2T)\cdot(2+T)\cdot2\dots(2-(l-3)T)}{6\cdot(6+T)\cdot(6+2T)\dots(6+(l-1)T)} \times
\frac{(4-3T)\cdot(4-2T)\dots(4+(l-4)T)}{T\cdot(2T)\dots(lT)}\times \\
\frac{4\cdot(4+(2l-3)T)\cdot(lT)\cdot(4-(3-l)T)}{(3T-4)^2\cdot(3T+4)}
\end{multline}

\subsubsection{\texorpdfstring{$l+k=3$}{Lg} and \texorpdfstring{$k=0$}{Lg}}
\begin{equation}
Z(x,0,3,T,4,-2)=-1
\end{equation}

\subsubsection{\texorpdfstring{$l+k=3$}{Lg} and \texorpdfstring{$l=0$}{Lg}}
\begin{multline}
Z(x,3,0,T,4,-2)=\\
=
\sinh\left[\frac{x}{4}: \right.
\frac{(2\alpha+\beta)\cdot6\cdot (6+T)\cdot(6+2T)\cdot(4+T)\cdot(4+2T)\cdot(4-3T)}{2\cdot4\cdot(3T)\cdot(2+T)\cdot(2+2T)\cdot(4-T)\cdot(4-2T)}
\end{multline}

\subsubsection{\texorpdfstring{$l+k=2$}{Lg} and \texorpdfstring{$k,l>0$}{Lg}}
\begin{multline}
Z(x,1,1,T,4,-2)=\\
=
-\sinh\left[\frac{x}{4}: \right.
\frac{(2\alpha+\beta)\cdot(2T)\cdot(6+2T)\cdot(6+T)\cdot(4+2T)\cdot(4+T)^2\cdot(8-2T)\cdot(2+2T)\cdot(4-T)}{2\cdot4^3\cdot6\cdot(T)^2(8-T)\cdot(2+T)\cdot(4-2T)
} 
\end{multline}

\subsubsection{\texorpdfstring{$l+k=2$}{Lg} and \texorpdfstring{$k=0$}{Lg}}
\begin{multline}
Z(x,0,2,T,4,-2)=
\sinh\left[\frac{x}{4}: \right.
=\frac{(4+T)\cdot(2+2T)\cdot(4+2T)\cdot(6+2T)\cdot(8-2T)}{2\cdot4\cdot6\cdot8\cdot(4-T)}
\end{multline}

\subsubsection{\texorpdfstring{$l+k=2$}{Lg} and \texorpdfstring{$l=0$}{Lg}}
\begin{equation}
Z(x,2,0,T,4,-2)=
\sinh\left[\frac{x}{4}: \right.
\frac{(6+T)\cdot(6+2T)\cdot (4+2T)}{2\cdot4\cdot(2+T)}
\end{equation}

\subsubsection{\texorpdfstring{$l+k=1$}{Lg} and \texorpdfstring{$l=0$}{Lg}}
\begin{equation}
Z(x,1,0,T,4,-2)=
\sinh\left[\frac{x}{4}: \right.
\frac{(2T)\cdot(6+2T)\cdot (4+T)}{2\cdot4\cdot(T)}
\end{equation}

\subsubsection{\texorpdfstring{$l+k=1$}{Lg} and \texorpdfstring{$k=0$}{Lg}}
\begin{multline}
Z(x,0,1,T,4,-2)=\\
=
\sinh\left[\frac{x}{4}: \right.
\frac{(2+2T)\cdot(4+2T)\cdot(6+2T)\cdot (8-2T)\cdot(4+T)}{2\cdot4\cdot6\cdot8\cdot(4-T)}
\end{multline}

{\bf Exceptional algebras:} 

For the exceptional algebras the procedure is technically similar to that for the classical algebras, so we omit its detailed presentation.

\end{document}